\documentclass[proof]{WileyASNA-v1}

\articletype{Article}

\received{}
\revised{}
\accepted{}

\begin{document}

\title{The role of quark matter surface tension in magnetars}

\author[1,2]{A.G. Grunfeld*}
\author[3]{G. Lugones}

\address[1]{\orgdiv{CONICET},\orgaddress{\state{Godoy Cruz 2290, (C1425FQB) Buenos Aires}, \country{Argentina}} }  

\address[2]{\orgdiv{Department of Theoretical Physics}, \orgname{CNEA}, \orgaddress{\state{Av. Libertador 8250, (1429) Buenos Aires}, \country{Argentina}}}

\address[3]{\orgdiv{Centro de Ci\^{e}ncias Naturais e Humanas}, \orgname{Universidade Federal do ABC}, \orgaddress{\state{Av. dos Estados 5001, CEP 09210-580, Santo Andr\'{e}, SP}, \country{Brazil}}}

\corres{*Av. Libertador 8250, (1429) Buenos Aires.  \email{grunfeld@tandar.cnea.gov.ar}}

\abstract{In spite of its key role in compact star physics, the surface tension of quark matter is not well comprehended yet.  In this work we analyze the behavior of the surface tension of three-flavor quark matter in the outer and inner core of cold deleptonized magnetars, proto magnetars born in core collapse supernovae, and  hot magnetars produced in binary neutron stars mergers. We explore the role of temperature, baryon number density, trapped neutrinos, droplet size, and magnetic fields within the multiple reflection expansion formalism.  Quark matter is described within the MIT bag model and is assumed to be in chemical equilibrium under weak interactions. We discuss some astrophysical consequences of our results.}


\maketitle

\section{Introduction}

Magnetars are apparently isolated compact objects characterized by a variable X-ray activity, whose main properties can be explained by the presence of extremely strong magnetic fields.
Their measured spin-down is related to a dipolar surface magnetic field that may be as high as up to $10^{15}$~G \citep{Kaspi}.  In general, it is believed that at least $10 \%$ of the young neutron-star population experience a magnetar phase during a short period of their evolution ($\sim 10^4$~years). During their lives they have intense activity, including persistent X-ray emissions, short bursts, large outbursts and giant flares which are believed to be powered by the decay of huge magnetic fields which drive changes in the crust  and the magnetosphere \citep{Duncan}.  Historically, magnetars have been associated with soft gamma repeaters (SGRs) and anomalous X-ray pulsars (AXPs), but some low magnetic field SGRs have also been found \citep{Rea1,Rea2}, as well as high surface $B$-field objects that behave as rotation powered pulsars with some occasional magnetar-like outbursts (see \citet{Archibald} and references therein).

The coincident observation of  GRB170817A  and  GW170817 confirmed the association between mergers of neutron stars and  short GRBs \citep{TheLIGOScientific:2017qsa,Monitor:2017mdv,Goldstein:2017mmi,Zhang:2017lpb}.  Short GRBs are often followed by an extended emission in lower energy electromagnetic bands referred to as an afterglow. The origin of the afterglow, particularly its X-ray emission, is still under discussion. Some models explain the X-ray afterglow plateau emission as being the result of the energy injection from a millisecond magnetar \citep{Dai:1998hm,Zhang:2000wx,Stratta:2018xza}.
Recently, an X-ray transient (CDF-S XT2) associated with a galaxy at redshift $z = 0.738$ has been detected whose light curve is analogous to the X-ray plateau of a GRB afterglow \citep{Xue:2019nlf}. Several features of CDF-S XT2 imply that this X-ray transient may have originated from a binary NS merger, and that the merger product is a long-lived millisecond magnetar \citep{Xue:2019nlf,Ren:2020eqd}.

The strong magnetic field of magnetars is thought to be due to amplification by a turbulent dynamo, which could be triggered by the magnetorotational instability \citep{Guilet:2015loa,Mosta:2015ucs} or by convection \citep{Thompson:1993hn,Raynaud:2020ist} during the first seconds following the core collapse of a massive star or a binary neutron star merger \citep{Giacomazzo:2014qba}. 
The resulting magnetic field configuration is expected to have both poloidal and toroidal components. Present numerical simulations suggest that dynamos generate a large-scale dipole magnetic field  reaching up to $10^{15} \, \mathrm{G}$ and that the toroidal component is always stronger than the dipole one  reaching values as high as $10^{16}\, \mathrm{G}$ \citep{Raynaud:2020ist}. These results, together with studies pointing that the small-scale magnetic field may be significantly larger than the large scale one \citep{Gourgouliatos:2016fnl} suggest that much larger magnetic field strengths could be found deep in the stellar core.

The internal composition of magnetars is not yet fully understood since  baryon number densities inside them may be as large as several times the nuclear saturation density, $n_0 \approx 0.16$~fm$^{-3}$ and first principle calculations are not available in this density regime.  In spite of this, it is widely accepted that deconfined quark matter would be present in neutron star cores. In this context, it is important to know the value of the surface tension $\sigma$ of quark matter because it is essential to understand whether the hadron-quark interface is simply a sharp discontinuity or a mixed phase where quarks and hadrons form geometrical structures that coexist over a wide density region of the star. If the energy cost of surface effects does not exceed the gain in bulk energy, the scenario involving a mixed phase turns out to be favorable \citep{Voskresensky:2002hu, Endo:2011em, Wu:2018zoe}. Surface tension also plays a crucial role in quark matter nucleation during the formation of compact stellar objects, because it influences the nucleation rate and the associated critical size of the nucleated drops \citep{Lugones:2011xv,Carmo:2013fr}.  It also determines the internal structure of strange stars which may fragment into a charge-separated mixture, involving positively-charged strangelets immersed in a negatively charged sea of electrons (see \citet{Lugones:2020lal}  and references therein). 

In this work we study the surface tension of three-flavor quark matter in cold magnetars as well as in hot magnetars with trapped neutrinos.  We extend previous works on this issue \citep{Lugones:2018qgu} by exploring the behaviour of $\sigma$ for a typical density of the outer core of magnetars ($n_B = n_0$) and comparing with a representative density of the inner core ($n_B = 4 n_0$). We will also focus on drop sizes close to the Debye screening length $\lambda_D$ in quark matter, which are more realistic than the large values explored previously \citep{Lugones:2018qgu}. We will show that the surface tension at the outer core is extremely  anisotropic for strong enough magnetic fields and we will explore some potential astrophysical consequences.  

The paper is organised as follows. In Section \ref{sec:MRE} we summarise the multiple reflection expansion (MRE)  formalism used to calculate the surface tension and in Section \ref{sec:astro} we describe the astrophysical scenarios where $\sigma$ will be determined.  In Section \ref{sec:results}  we show our results and  in Section \ref{sec:conclusions}  we present our conclusions.

%---------------------------------------------------
\section{The formalism}

\subsection{Surface tension in the MRE framework}
\label{sec:MRE}

In the present study, we use the MIT bag model \citep{Glendenning:1997wn} to describe  small drops of quark matter composed by $u,d,s$ quarks and electrons in chemical equilibrium under weak interactions, assuming  local electric charge neutrality. Finite size effects are included within the MRE formalism assuming that a finite  droplet with an arbitrary shape has a modified  density of states $\rho_\mathrm{MRE}$ given by \citep{Balian:1970fw}:
\begin{equation}
\rho_\mathrm{MRE}(k,m_i,S, V, \cdots) = 1 + \frac{2 \pi^2}{k} \frac{S}{V} f_{S,i}  + \cdots
\label{rho_MRE}
 \end{equation}
where $m_i$ is the mass of the particle, $S$ is the droplet's surface, $V$ its volume, and 
\begin{equation}
f_{S,i}(k) = - \frac{1}{8 \pi} \left(1 - \frac{2}{\pi} \arctan \frac{k}{m_i} \right) .
\label{eq:fs}
\end{equation}
To implement the MRE formalism in a generic thermodynamic integral $I$  we have to perform the following replacement:
\begin{equation}
I \equiv \frac{1}{(2 \pi)^3} \int \cdots d^3 k \rightarrow    \frac{1}{(2 \pi)^3}  \int_{\Lambda }^{\infty} \cdots \rho_{\mathrm{MRE}}(k) \,   4 \pi k^2 dk.
\label{MRE}
\end{equation}

For matter immersed in a magnetic field $\textbf{B}$ pointing in the $z$ direction, the transverse motion of particles with electric charge $qe$ is quantized into Landau levels.
The  momentum $k$ is given by  $k =  \sqrt{k_z^2 +  2 \nu |q e B| }$   where $\nu \geq 0$ is an integer, and the momentum integrals  in the transverse plane must be replaced by sums over the discretized levels. Thus, the thermodynamic integral in Eq. (\ref{MRE}), now reads \citep{Lugones:2016ytl}
\begin{equation}
I \rightarrow     \frac {|q e B|}{2 \pi^2} \sum_{\nu=0}^{\infty} \alpha_{\nu}  
\int_{\Lambda_{\nu}}^{\infty} \cdots \rho_{\mathrm{MRE}} dk_z ,
\label{MRE_with_B}
\end{equation}
where $\alpha_{\nu} =2$ for all cases except for ${\nu} = 0$, where $\alpha_{\nu} =1$. The infrared cutoff  $\Lambda_{\nu}$ in the momentum along the direction of the magnetic field is the largest solution  of  the equation  $\rho_{\mathrm{MRE}}(k_z, m_i,  S, V) = 0$ with respect to the momentum $k_z$.  Taking only the first two terms of Eq. (\ref{rho_MRE}) we obtain \citep{Lugones:2016ytl}:
\begin{equation}
\Lambda_{i,\nu}  =  \sqrt{ \frac{S^2}{4V^2} x^2_{0,i}  -   2 \nu |q_i e B| } ,
\label{LIR}
\end{equation}
where $x_{0,i}$ is the solution of  $\lambda x = \cot x$  with $\lambda = S / (2Vm_i)$.  

 The surface tension in the directions parallel and transverse to the magnetic field are \citep{Lugones:2018qgu}:
\begin{eqnarray}
\sigma_i^{\parallel} &=& - |q_i e B| \sum_{\nu=0}^{\infty} \alpha_{\nu} 
\int_{\Lambda_{i,\nu}}^{\infty}   \frac { (F_i + {\bar F}_i)  f_{S,i}  k_z^2  dk_z}{k \sqrt{k^2 + m_i^2}},
\end{eqnarray}
\begin{eqnarray}
\sigma_i^{\perp} & = & - |q_i e B|^2 \sum_{\nu=0}^{\infty} \alpha_{\nu} \nu  \int_{\Lambda_{i,\nu}}^{\infty}  \frac { (F_i + {\bar F}_i) f_{S,i} \, dk_z}{k  \sqrt{k^2 + m_i^2}},
\end{eqnarray}
where the Fermi-Dirac distribution functions for particles and antiparticles are respectively: 
\begin{eqnarray}
F_i &=& \frac{1}{e^{(E_i - \mu_i)/T} + 1} , \\
{\bar F}_i &=& \frac{1}{e^{(E_i + \mu_i)/T} + 1} ,
\end{eqnarray}
being $E_i = (k_z^2  + 2 \nu |q_i e B| + m_i^2)^{1/2}$ and $\mu_i$ the chemical potential for the particle species $i$. 

The total parallel/transverse surface tension reads
\begin{equation}
\sigma_T^{\parallel, \perp} = \sum_{i = u, d, s, e} \sigma_i^{\parallel,\perp}.    
\end{equation}
Within the MRE formalism, neutrinos have vanishing surface tension because they are assumed to be massless. However they contribute indirectly to $\sigma_T^{\parallel, \perp}$ because they shift the chemical equilibrium conditions affecting the relative abundances of different particle species.

\subsection{Astrophysical scenarios}
\label{sec:astro}

We focus here on droplets of quark matter in equilibrium under weak interactions. Thus, the chemical potentials are related by:
\begin{eqnarray}
\mu_d &=& \mu_u + \mu_e - \mu_{\nu_e} ,  \\
\mu_s &=& \mu_d. 
\end{eqnarray}
We also assume that charge neutrality holds locally, which means that:
\begin{eqnarray}
\tfrac{2}{3} n_u - \tfrac{1}{3} n_d - \tfrac{1}{3} n_s - n_e = 0, 
\label{charge_neutrality}
\end{eqnarray}
where the number densities are given by (\cite{Lugones:2018qgu})
\begin{equation}
n_i  =  \frac{|q_i e B|}{2\pi^2}  \sum_{\nu=0}^{\infty} \alpha_{\nu}   \int_{\Lambda_{i,\nu}}^{\infty} 
(F_i - {\bar F}_i) \left[1 + \frac{2 \pi^2 S}{k V} f_{S,i}\right]  dk_z.
\end{equation}

In the present study we will concentrate on two different values for the baryon number density $n_B$. The lowest one, $n_B = n_0$, is a representative density of the outer core of a compact star, and the second one, $n_B = 4 n_0$, of the inner core. Regarding the magnetic field strength, we consider a low value and a high one: $eB_\mathrm{low} = 5  \times  10^{-3} \, \mathrm{GeV}^2 = 8.5 \times 10^{17} \, \mathrm{G}$ and $eB_\mathrm{high} = 5  \times 10^{-2}\, \mathrm{GeV}^2 = 8.5 \times 10^{18} \, \mathrm{G}$.

We will analyse three different astrophysical scenarios according to the typical temperature and the role of neutrinos: 
\begin{itemize}

\item \textit{Cold magnetars (CM)}. The thermodynamic state is characterized by a very low temperature and neutrino transparency. We shall consider here $ 1 \, \mathrm{MeV} < T < 10 \, \mathrm{MeV}$  and $\mu_{\nu_e}=0$.   
    
\item \textit{Proto magnetars (PM)}. Matter is non-degenerate and there is a considerable amount of trapped neutrinos. As a representative case we adopt $10 \, \mathrm{MeV} < T < 50 \, \mathrm{MeV}$ and $\mu_{\nu_e} = 100 \, \mathrm{MeV}$. 
    
\item \textit{Post merger magnetar (PMM)}. According to numerical simulations \citep{Most:2018eaw}, a compact object produced in a neutron star merger may attain temperatures of several tens of $\mathrm{MeV}$  and contains a large amount of trapped neutrinos. We consider here $50  \, \mathrm{MeV} < T < 100$ MeV and the extreme value $\mu_{\nu_e} = 200$ MeV. 
    
\end{itemize}

\section{Results}
\label{sec:results}

%%%%%%%%%%%%%     TABLE 1    %%%%%%%%%%%%% 
\begin{table}[tb]
\centering
\begin{tabular}{c c c | c}
\hline \hline
particles & $m$ [MeV]          &          $V/S$ [fm]              &   $x_0$               \\
\hline  
electrons &0.511     &    2   &   0.101601 \\
&0.511     &   10                 &  0.225631  \\   
 \hline 
quarks u, d &5        &   2    &  0.313081   \\  
&5 &   10                 &  0.657008  \\ 
\hline
quarks s &150   &   2        &  1.19603    \\ 
&150     &   10              &  1.47413  \\ 
\hline \hline 
\end{tabular}
\caption{Solution of $\lambda x_0 = \cot x_0$  with $\lambda = S / (2Vm)$ for  different particle masses and two different values of $V/S$.} \label{table:cutoff}
\end{table}

\begin{figure*}[tb]
\centering
\includegraphics[angle=0,scale=0.35]{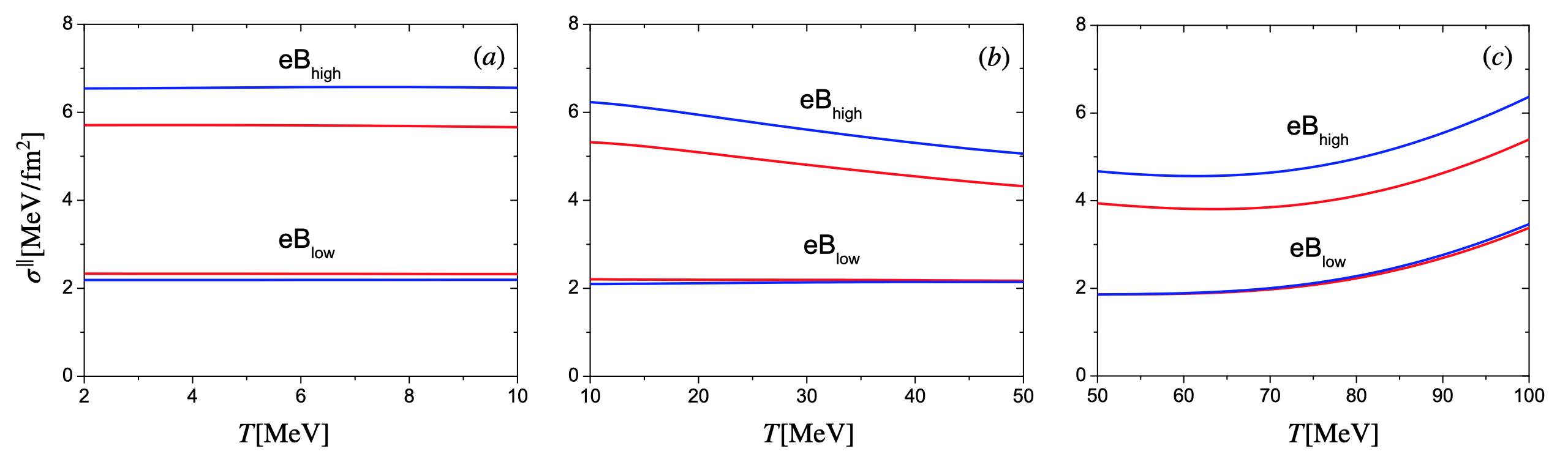} 
\includegraphics[angle=0,scale=0.35]{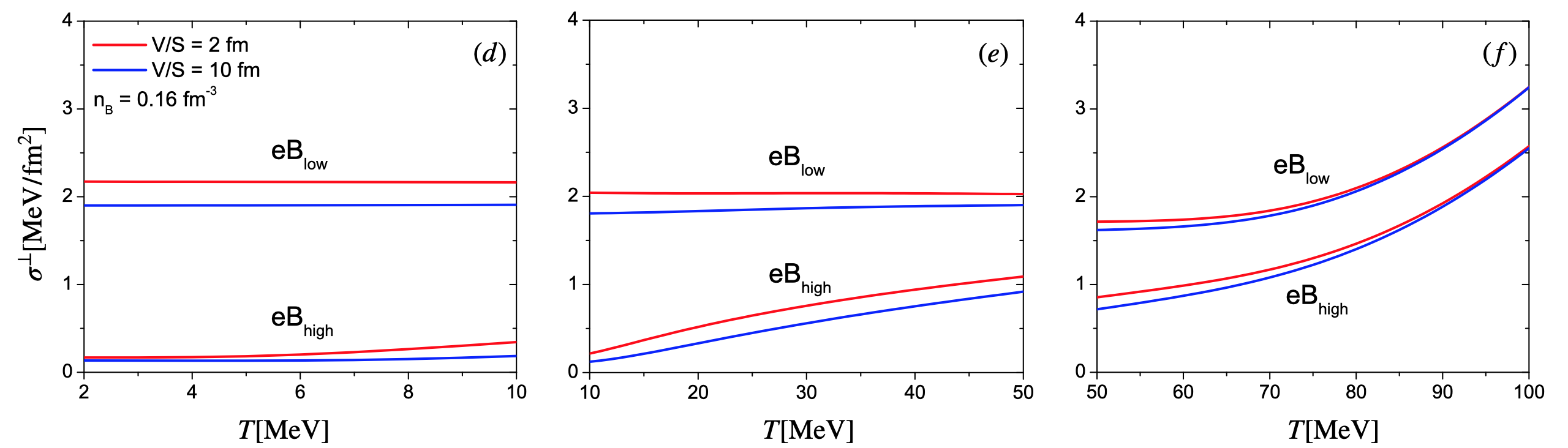} 
\caption{Parallel surface tension (upper panels) and transverse surface tension (lower panels) for $n_B = n_0$. Panels (a) and (d) correspond to CMs, panels (b) and (e) to PMs and panels (c) and (f) to PMMs. } 
\label{fig1}
\end{figure*}

\begin{figure*}[tb]
\centering
\includegraphics[angle=0,scale=0.35]{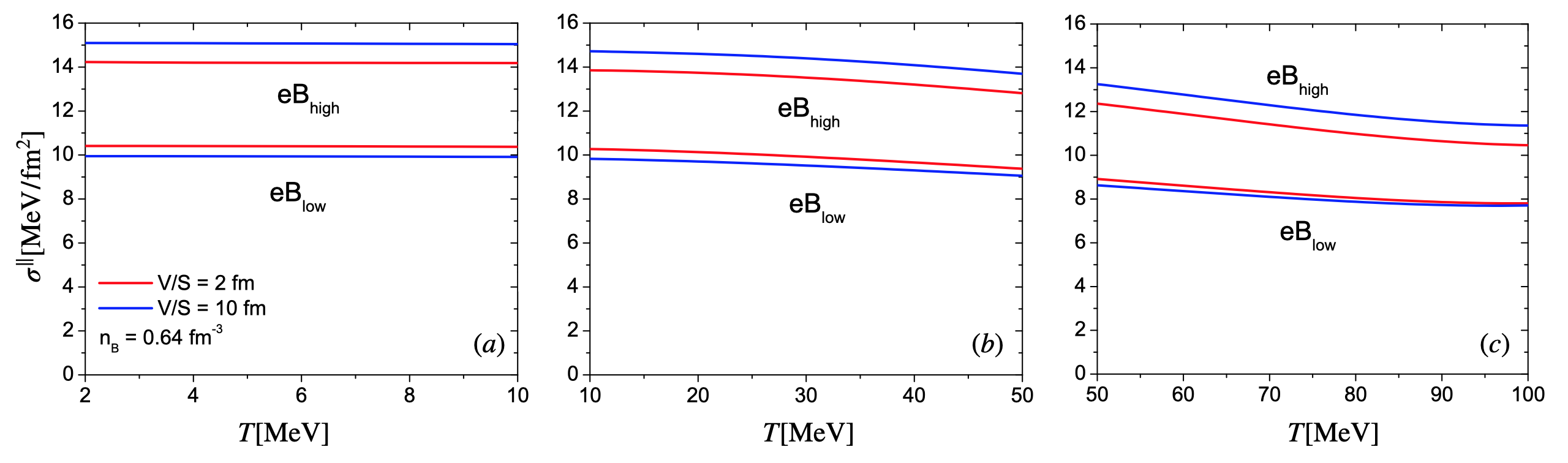} 
\includegraphics[angle=0,scale=0.35]{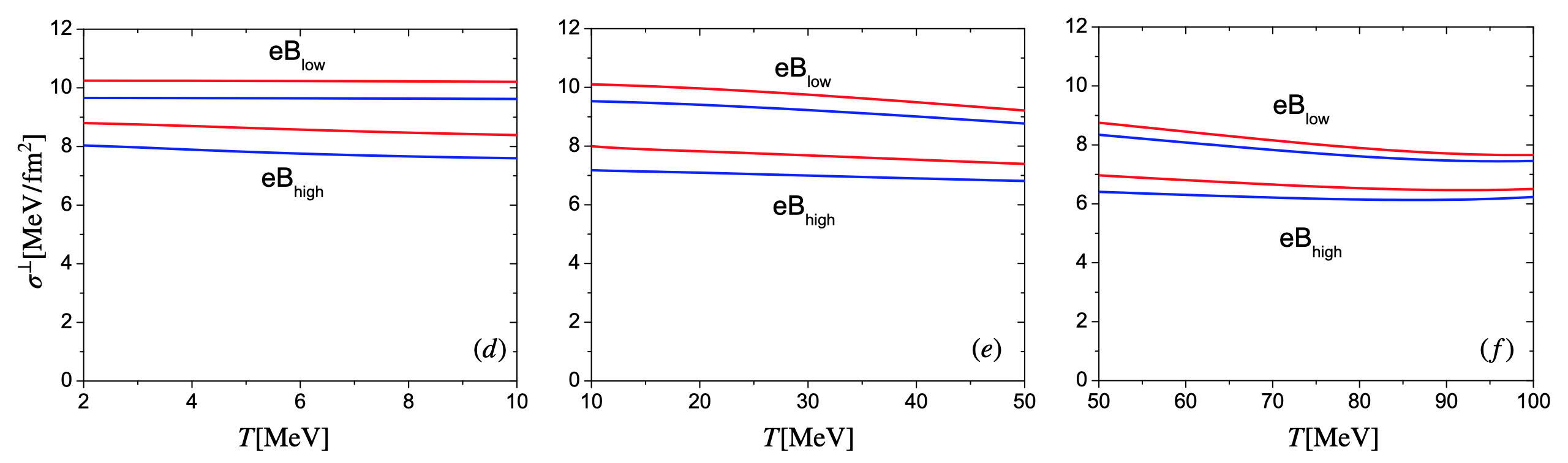} 
\caption{Same as in Fig. \ref{fig1} but for $n_B = 4 n_0$.} 
\label{fig2}
\end{figure*}

In the present work we consider $m_u = m_d =$ 5 MeV, $m_s$ = 150 MeV and $m_e$ = 0.511 MeV. We adopt two different drop sizes close to the Debye screening length in quark matter ($\lambda_D \sim 5 \; \mathrm{fm}$): $V/S = 2 \; \mathrm{fm}$ and $V/S = 10 \; \mathrm{fm}$. In Table \ref{table:cutoff} we show the corresponding values for $x_{0}$ that allow determining the infrared cutoff $\Lambda_{\nu}$. As mentioned before, we consider two values for the magnetic field, namely $eB_\mathrm{low}$ and $eB_\mathrm{high}$. For magnetic field strengths smaller than $eB_\mathrm{low}$, we already know that the results for $\sigma$ coincide with those for $eB_\mathrm{low}$ \citep{Lugones:2016ytl, Lugones:2018qgu}. 

In Figs. \ref{fig1} and \ref{fig2} we summarize our results. Both figures display the values for $\sigma^{\parallel}$ and $\sigma^{\perp}$ as a function of $T$, taking as inputs the baryon number density $n_B$, the magnetic field $eB$ and the size of the system $V/S$. The three different horizontal panels correspond to each astrophysical scenario studied in this work - CM, PM and PMM - characterized by the range of temperature and the amount of neutrinos trapped in the system. Fig. \ref{fig1} shows our results for $n_B = n_0$, which is a typical density of the outer core of a compact star, and Fig. \ref{fig2} for $n_B = 4 n_0$, a characteristic density in the inner core. These choices complement our previous studies \citep{Lugones:2018qgu}, by considering both smaller quark drops (closer to  $\lambda_D$ in quark matter) and a lower density, $n_B = n_0$.

As a general feature we observe that the surface tension increases with the density, reaching values as large as $15 \; \mathrm{MeV}/\mathrm{fm}^2$ at low $T$ and $n_B = 4 n_0$. As seen in both figures, when the magnetic field strength increases, the value of $\sigma^{\parallel}$ gets larger, but $\sigma^{\perp}$ is reduced. 
The size of the system,  $V/S$, affects the surface tension (both longitudinal and transverse). This effect is more evident for $eB_\mathrm{high}$ and low densities, with relative differences of $\sim 10\%$ between the results for $V/S=2 \, \mathrm{fm}$ and $V/S=10 \, \mathrm{fm}$.

For cold magnetars, the surface tension is insensitive to the temperature as seen in panels (a) and (d) of Figs. \ref{fig1} and \ref{fig2}.
For PMs, the behaviour of $\sigma$ depends on the magnetic field strength and the density considered. For $n_B = n_0$ and $eB_\mathrm{low}$, both $\sigma^{\parallel}$ and $\sigma^{\perp}$  are insensitive to the temperature (see panels (b) and (e) of Fig.  \ref{fig1}). However, for $eB_\mathrm{high}$,  $\sigma^{\parallel}$ decreases  and  $\sigma^{\perp}$ grows with $T$. For $n_B = 4 n_0$, (panels (b) and (e) of Fig. \ref{fig2}) there is a slight decrease of $\sigma$ with $T$, for both magnetic field strengths considered here. 
Finally, for PMMs, the surface tension increases with $T$ for low densities (panels (c) and (f) of Fig. \ref{fig1}), but slightly decreases with $T$ for higher densities (panels (c) and (f) of Fig. \ref{fig2}).

\section{Conclusions}
\label{sec:conclusions}

In the present work we have studied the surface tension of quark matter droplets in magnetars. We assumed that quark matter is a locally charge neutral mixture of $u, d, s$ quarks,  electrons and neutrinos, all in chemical equilibrium under weak interactions and immersed in a strong magnetic field. The equation of state was described by the MIT bag model and finite size effects were included by means of the MRE formalism. The presence of a strong magnetic field quantises the transverse motion of charged particles  into Landau levels resulting in a surface tension that has a different value in the parallel and the transverse directions with respect to the magnetic field. As input parameters we have picked two values of the magnetic field, $eB_\mathrm{low} = 5  \times  10^{-3} \, \mathrm{GeV}^2$ and $eB_\mathrm{high} = 5  \times 10^{-2}\, \mathrm{GeV}^2$ and two different sizes of the quark matter droplets, $V/S = 2$ fm and $10$ fm. For the baryon number density we considered $n_B = n_0$ which is a typical value for the outer core and $n_B = 4n_0$ which is representative of the inner core of a magnetar. We focused our analysis on three different astrophysical scenarios (CM, PM and PMM), according to their temperature and the amount of trapped neutrinos.

Our results show that, for the densities considered here, the surface tension spans values between $0.2 \; \mathrm{MeV} / \mathrm{fm}^{2}$  and $15 \; \mathrm{MeV} / \mathrm{fm}^{2}$.
We also find that, for magnetic fields smaller than $eB_\mathrm{low}$,   there is no difference between the transverse and the longitudinal values of the surface tension, meaning that quark drops must have a spherical shape. However, for larger magnetic fields the longitudinal contribution is larger than the transverse one and elongated shapes in the direction of the magnetic field are more prone to occur. This effect is more pronounced at low temperatures.  

Surface tension has a key role in several astrophysical contexts.  It plays a crucial role in quark matter nucleation during the formation of compact stellar objects, because it determines the nucleation rate and the associated critical size of the nucleated drops \citep{Carmo:2013fr,Lugones:2011xv}. Small values of the surface tension, as the ones obtained here, tend to favor faster nucleation timescales than other models such as NJL \citep{Lugones:2011xv}.

Finite size effects are also determinant in the formation of mixed phases at the core of hybrid stars which may arise only if the surface tension is smaller than a critical value $\sigma_\mathrm{crit} \approx 60 \; \mathrm{MeV} / \mathrm{fm}^{2}$ \citep{Voskresensky:2002hu, Endo:2011em, Wu:2018zoe}. Below this value, the structure of the mixed phase becomes mechanically unstable and local charge neutrality is recovered.  Our results are significantly smaller than $\sigma_\mathrm{crit}$ indicating that a mixed phase would be favoured. Although our calculations assumed that matter is locally electrically neutral, we don't expect that a more realistic calculation assuming global charge neutrality will change the order of magnitude of $\sigma$, and therefore the above conclusion would still be valid. However, since large magnetic fields lead to different values for $\sigma^{\parallel}$ and $\sigma^{\perp}$, the pasta configurations that may appear in the mixed phase (droplets, rods, slabs, tubes, and bubbles) may be different in non-magnetized hybrid stars and in hybrid magnetars. 

Finally, the surface tension may affect decisively the internal structure of self-bound strange stars which may fragment into a charge-separated mixture, involving positively-charged strangelets immersed in a negatively charged sea of electrons, presumably forming crystalline solid matter \citep{Jaikumar:2005ne}. This would happen below another critical surface tension $\sigma^{\prime}_\mathrm{crit}$ whose value depends strongly on the equation of state.  Present estimations of $\sigma^{\prime}_\mathrm{crit}$ give values in the range $0.5 - 18\; \mathrm{MeV}/\mathrm{fm}^2$ \citep{Alford:2008ge, Alford:2006bx}. Assuming global charge neutrality and  non-magnetized matter, we have shown in a recent work \citep{Lugones:2020lal} that this phase is favored in a scenario of high enough $\sigma^{\prime}_\mathrm{crit}$ (above $\sim  10 \; \mathrm{MeV}/\mathrm{fm}^2$). However, we have shown here that for strongly magnetized matter, $\sigma^{\parallel}$ grows and $\sigma^{\perp}$ diminishes with respect to the non-magnetized case.  This leads to a much more involved situation because a single critical surface tension is not enough to determine whether a charge-separated mixture is favoured. A more detailed analysis would be needed to determine the true ground state of self-bound ultra-magnetized matter.

\section*{acknowledgments}
A. G. G would like to acknowledge to CONICET for financial support under Grant No. PIP17-700. G.L. is thankful to the Brazilian agency Conselho Nacional de Desenvolvimento Cient\'{\i}fico e Tecnol\'ogico (CNPq) for financial support.

\bibliography{IWARA}

\end{document}